\documentclass[preprint,nocite]{epl}
\usepackage{graphicx,amsmath,amssymb}
\newcommand{\bb}{\begin{equation}}
\newcommand{\en}{\end{equation}}
\renewcommand{\vec}[1]{{\mathbf #1}}

\title{Dynamics of active membranes with internal noise}
\author{D.\ Lacoste \inst{1} \and A.W.C.\ Lau \inst{2,3}
}
\institute{\inst{1} Laboratoire de Physico-Chimie Th\'eorique, UMR
7083 CNRS-ESPCI, 10 rue Vauquelin, F-75231 Paris cedex 05, France\\
\inst{2} Department of Physics and Astronomy, University of
Pennsylvania, Philadelphia, PA 19104, USA \\
\inst{3} Department of Physics, Florida Atlantic University, Boca Raton, FL 33431, USA }

\pacs{87.16.-b}{Subcellular structure and processes}
\pacs{05.40.-a}{Fluctuation phenomena, random processes, noise, and Brownian motion}
\pacs{05.70.Np}{Interface and surface thermodynamics}

\begin{document}

\maketitle

\begin{abstract}
We study the time-dependent height fluctuations of an active
membrane containing energy-dissipating pumps that drive the
membrane out of equilibrium. Unlike previous investigations based
on models that neglect either curvature couplings or random
fluctuations in pump activities, our formulation explores two new
models that take both of these effects into account.  In the first
model, the magnitude of the nonequilibrium forces generated by the
pumps is allowed to fluctuate temporally. In the second model, the
pumps are allowed to switch between ``on" and ``off" states. We compute
the mean squared displacement of a membrane point for both models,
and show that they exhibit distinct dynamical behaviors from previous models, and
in particular, a superdiffusive regime specifically arising from the shot noise.
\end{abstract}

\section{Introduction}

While the physics of biomembranes in equilibrium is fairly
developed \cite{seifert}, recent studies focus on active membranes
that contain proteins, such as ion channels, ion pumps, and
photo-active proteins like bacteriorhodopsin. These proteins
consume the chemical energy of ATP, dissipate it into the medium,
and thus drive the membrane out of equilibrium
\cite{cell,ramaswamy2}. The importance of active processes has
been demonstrated in an experiment showing that the fluctuations
in the shape of red blood cells depend on the viscosity of the
environment and on ATP concentrations \cite{tuvia}.  In {\em in
vitro} experiments, nonequilibrium forces arising from ion pumps
embedded in a membrane are shown to enhance its fluctuations
\cite{bassereau1,bassereau2,girard}. There are currently two
theoretical models for active membranes \cite{ramaswamy2}. The
Prost-Bruinsma (PB) model takes nonequilibrium forces in the form
of active noises that include diffusion and the stochastic nature
(shot noise) of the pumps, but ignores the coupling between the
pumps and membrane curvature \cite{prost1,prost2,Granek}.   The
other model proposed by Ramaswamy, Toner and Prost (RTP)
incorporates this coupling but ignores the random nature of the
protein activity \cite{ramaswamy3,sankararaman}.  For {\em steady
state} measurements of active membranes, the RTP model agrees quite well with
experiments \cite{bassereau2,girard}.  In this Letter, we further
explore the dynamical properties of the RTP model, argue that it
is important to include the shot noise for {\em dynamical}
measurements, and present two new models that include both
curvature effects and pump stochasticity. In the first model,
which may be an appropriate description for light-activated pumps
such as bacteriorhodopsin, the magnitude of the nonequilibrium
force fluctuates on a time scale that is fast compared to that of
membrane fluctuations.  The second model, the two-state model,
which may be realized in typical ion channels \cite{cell},
describes pumps that are able to switch from ``on" to ``off" state
on a time scale that is long compared to membrane fluctuation time.
These models represent natural generalizations of the RTP model, and they show distinct
dynamical behaviors. In particular, the two-state model predicts
that the mean-squared-displacement (MSD) of a membrane point
exhibits superdiffusion in the experimentally relevant regime,
whereas the RTP model would predict subdiffusion. Our predictions
should be accessible to microrheological experiments as carried
out in Ref.\ \cite{Chatenay} for similar systems.

\section{Membrane dynamics in the RTP model}
\label{sec:background}

Consider a 2D tensionless, asymptotically flat, fluid membrane
embedded a 3D space, lying on the $x$-$y$ plane. Confined to move
in the membrane are two types of mobile active pumps.  They can
either be oriented up or down (for the orientation of the force
dipole see \cite{sankararaman}) with respect to the normal of the
membrane, whose shape is described by a height field $h({\bf r},
t)$ \cite{flat}. Their local coarse-grained densities are,
respectively, $\rho^{\uparrow}({\bf r}, t)$ and
$\rho^{\downarrow}({\bf r},t)$, where ${\bf r}$ denotes the 2D
position vector. The RTP model describes the dynamics of $h({\bf
r}, t)$ and the imbalance field, $\psi({\bf r}, t) \equiv
\rho^{\uparrow}({\bf r},t) - \rho^{\downarrow}({\bf r},t)$, taking
into account the curvature coupling and nonequilibrium forces
arising from the pump activities.  Here, we first summarize the
RTP model and further discuss its dynamical properties.

In the regime where inertial effects are negligible, the dynamics
of an active membrane is governed by Darcy's law, which states
that the relative flow of the solvent with respect to the membrane
is proportional to the normal force per unit area, $F_m({\bf
r},t)$, exerted on it: $\partial_t h({\bf r},t ) - v_{sz}({\bf
r},t)= \lambda_p  F_m({\bf r},t)$, where $v_{sz}({\bf r},t)$ is
the $z$-component of the fluid velocity field at the surface of
the membrane, $\lambda_p$ is the membrane permeability, and
$F_m({\bf r},t)$ is the sum of two contributions: a passive part,
$F_p({\bf r},t) = - { \delta {\cal H} / \delta h({\bf r},t) }$,
arising from membrane elasticity and an active part from pump
activities \cite{bassereau2,ramaswamy3}, $F_A({\bf r}, t) =
F_1({\bf r},t)\, \psi(\vec{r}\,t) + F_2({\bf r},t)\, \rho({\bf
r},t)\,\nabla^2 h({\bf r},t)$, where $\rho({\bf r},t) \equiv
\rho^{\uparrow}({\bf r},t)+ \rho^{\downarrow}({\bf r},t)$. Note
that the expression for $F_A({\bf r}, t)$ is the only possible
form that is linear in the density fields obeying the symmetry:
$(h,\psi, \rho, F_A) \rightarrow (-h,-\psi,\rho, - F_A)$.  In the
RTP model, $\rho({\bf r},t)= \rho_0$ is assumed to be uniform, and
more importantly, $F_1({\bf r},t) = F_1$ and $F_2({\bf r},t) =
F_2$ are assumed to be constant in space and time, capturing only
an average force produced by each pump. (For simplicity, we set
$F_2=0$ throughout this paper, {\em i.e.\ } ignoring higher order
contributions from the $F_2$ term.)  The membrane free energy is given by \cite{leibler}\bb
{\cal H} =\frac{1}{2} \int d^2{\bf r} \left[ \, \kappa ( \nabla^2 h )^2 +
\chi \psi^2 - 2\,\Xi\,\psi \nabla^2 h\,\right ],
\end{equation}
where $\kappa$ is the bare bending modulus, $\chi$ is the osmotic modulus of
the pumps, and $\Xi$ is the curvature coupling parameter which arises from
head-tail asymmetry of the pumps.  In equilibrium, $\kappa$ is renormalized to
$\kappa_e \equiv \kappa - \Xi^2/\chi$, which must be positive to ensure stability.
The fluid velocity $\vec{v}({\bf x}, t)$ surrounding the membrane
obeys Stokes' law: $ \eta \nabla^2 \vec{v}({\bf x}, t) = \nabla p({\bf x},t) -
\vec{F}_m({\bf x}, t)$, where $\eta$ is the solvent viscosity, $p({\bf x},t)$ is the
pressure field which ensures the incompressibility condition:
$\nabla \cdot \vec{v} = 0$, and $\vec{F}_m({\bf x}, t)$ is the total force
exerted on the fluid by the membrane.  It contains two parts:
a passive part, $\vec{F}_p({\bf x},t) = -
\left [\,{ \delta {\cal H} / \delta h({\bf r},t) }\right ]\delta(z)\,\hat{\bf z} $,
arising from membrane elasticity and an active part arising from the force exerted by
the pumps on the fluid, modelled as a dipolar force density with force centers
located at $z=w$ and $z= - w'$ \cite{bassereau2,sankararaman}:
$\vec{F}_A({\bf x}, t) = F_A({\bf r}, t) \left[\,\delta \left( z-w \right)
- \delta \left( z+w' \right)\,\right]\,\hat{\bf z}$.  Assuming $\psi(\vec{r} ,t)$
obeys the conserved dynamics: $\partial_t \psi = \Lambda \nabla^2 \delta {\cal H}/\delta \psi + \nu$
and eliminating the fluid velocity from Darcy's law,
we obtain in Fourier space two coupled Langevin equations \cite{bassereau2}\bb
\begin{array}{c}
\partial_t h({\bf q},t) + \omega_h h({\bf q},t) =  \beta\,\psi({\bf q},t) + \mu({\bf q},t),\\
\partial_t \psi({\bf q},t) + \omega_\psi \psi({\bf q}, t) = \gamma\,h({\bf q},t) + \nu({\bf q},t),
                    \end{array}
\label{Langevin1}
\en
where $\omega_h = \kappa q^3/( 4 \eta)+
\kappa \lambda_p  q^4$, $\beta = \lambda_p F_1 - (P_1 w  + \Xi)
q/(4 \eta) - \Xi\,\lambda_p q^2$, $\omega_\psi =
\Lambda\,\chi\,q^2 \equiv D q^2$, $D$ is the pump's diffusion
constant, $\gamma = -\Lambda\,\Xi\,q^4$, and
$P_1=F_1(w^2-w'^2)/(2w)$. The noises $\mu({\bf q},t)$ and
$\nu({\bf q},t)$ are assumed to be thermal in origin:
$\langle\,\mu({\bf q},t) \rangle =  \langle \nu({\bf q},t) \rangle
=0 $, and $\langle\,\mu({\bf q},t)\,\mu(-{\bf q},t') \rangle = 2
k_B T [\,\lambda_P + 1/(4 \eta q )\,] \delta(t-t') \equiv \Gamma_1
\delta(t-t')$ and $\langle\,\nu({\bf q},t)\,\nu(-{\bf q},t')
\rangle = 2 k_B T \Lambda q^2 \delta(t-t') \equiv \Gamma_2
\delta(t-t') $, where $\Lambda$ is the pump mobility,
$k_B$ is the Boltzmann constant and $T$ is the temperature.

It is straightforward to evaluate the two-point correlation function,
$\langle h({\bf q},t)h(-{\bf q},0) \rangle$:\bb
\langle h({\bf q},t)h(-{\bf q},0) \rangle   =
{\Gamma_1 \left[\,M_+\,e^{-\omega_+ t} - M_-\,e^{-\omega_- t}\,\right] \over
2 A B \left(\omega_+ -\omega_- \right)}
   + { \Gamma_2\,\beta^2 \left[\,\omega_+\,e^{-\omega_- t} -  \omega_-\,e^{-\omega_+ t}\,\right] \over
   2 A B \left(\omega_+ -\omega_- \right)},
\label{Uneq_time_corr}
\en
where $A \equiv \omega_h \omega_\psi-\beta \gamma$, $B \equiv \omega_h +
\omega_\psi$, $\omega_{\pm} = [ \, \omega_h + \omega_\psi
\pm \sqrt{(\omega_h-\omega_\psi)^2+4\beta \gamma}\, ]/2$,
and $M_{\pm} \equiv  A\, \omega_{\pm} - \omega_{\psi}^2\,\omega_{\mp}$.
We assume that $\omega_{\pm} > 0$, so that the system is
dynamically stable and reaches a steady state whose distribution
is Gaussian, as expected.  Its variance is given by [setting $t=0$ in Eq.\ (\ref{Uneq_time_corr})],
$\langle h({\bf q},0)h(-{\bf q},0) \rangle =
[\,\Gamma_1   (\,A+\omega_\psi^2\,) + \Gamma_2  \beta^2\,]/(2AB)
\sim k_B T_{\mbox{\scriptsize eff}}/(\kappa_e q^4)$, in the $\lambda_p \sim 0$ limit \cite{note1},
where $T_{\mbox{\scriptsize eff}} = \kappa_e T
[ 1 + P_1 w ( \Xi + P_1 w )/( \chi \kappa )]/( \kappa_e - P_1 w\,\Xi/\chi)$ \cite{bassereau2,girard}.
This effective temperature $T_{\mbox{\scriptsize eff}}$ is found to
be largely consistent with the experimental observations.

\begin{figure}
{\par {{\includegraphics[scale=0.35]{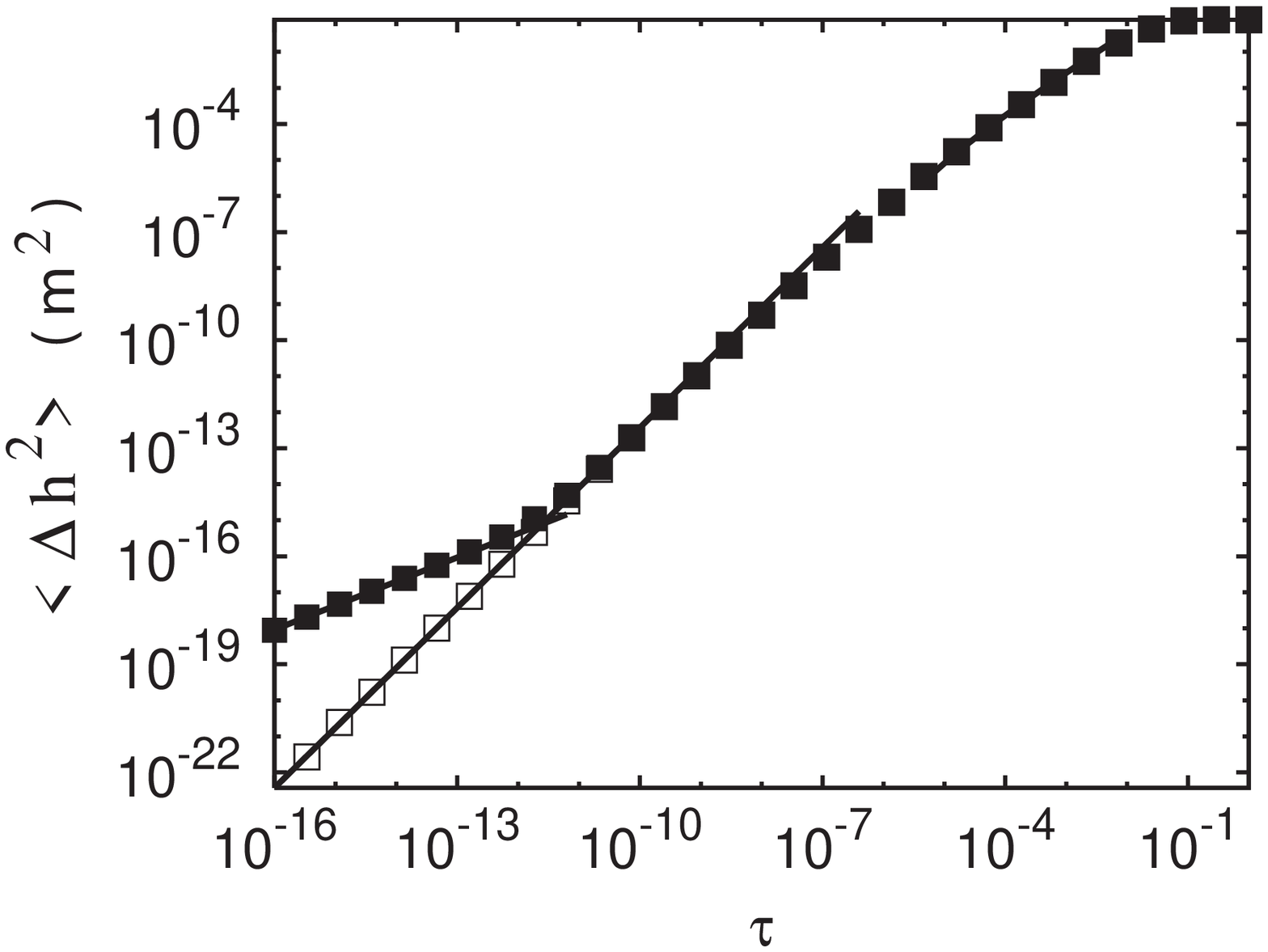}}
{\includegraphics[scale=0.35]{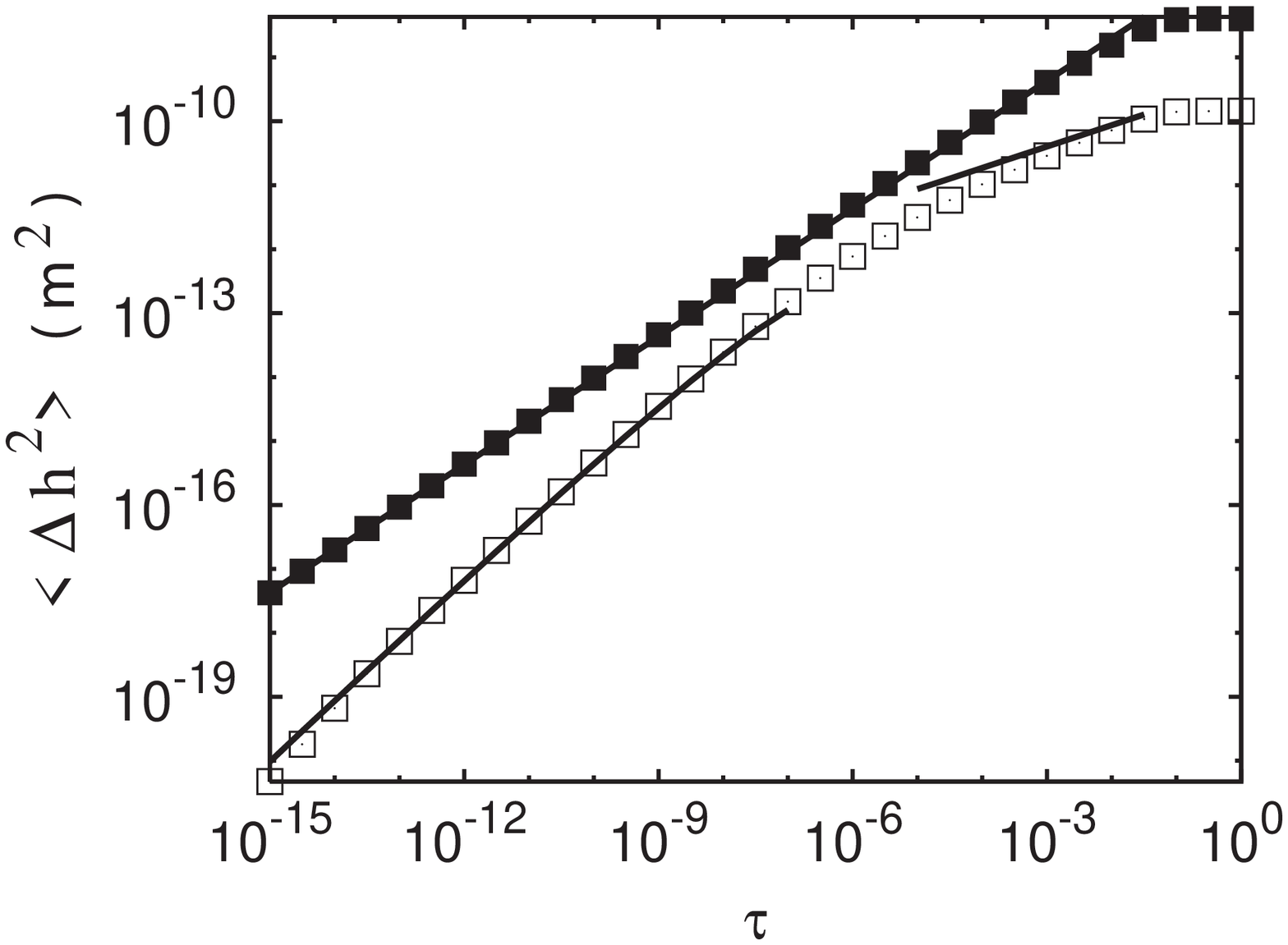}}}
\par}
\caption{The active part of the MSD $\langle \Delta h^2_a (t)
\rangle$ (empty symbols) and the overall MSD (full symbols)  as a
function of $\tau \equiv t/t_L$ for the RTP model in the
permeation dominated regime ($\lambda_p \neq 0$, Left) and the
regime with negligible permeation ($\lambda_p =0$, Right). The
solid lines are scaling laws of Eqs.~(\ref{Granek MSD}) and
(\ref{My MSD}) which compare well with points obtained from
numerically integrating Eq.\ (\ref{Uneq_time_corr}). The
parameters used are drawn from experimental values in Ref.\
\cite{bassereau2,girard}: $\kappa=10\,k_B T$, $\eta=10^{-3}\,$kg
m$^{-1}$s$^{-1}$, $w = 5\,$nm, $w' = 4\,$nm, $\Xi=w k_B T$, $P_1 =
10\,k_B T$, $D \sim 10^{-12}$m$^2$/s, $L=1\,$mm, $\rho_0 = 3 \cdot
10^{15}\,$m$^{-2}$, and when $\lambda_p \neq 0$ we choose $\beta=5
\cdot 10^{-18}\,$m$^3$/s. In both cases, the crossover time is
$t_{c1}/t_L \sim 10^{-6}$, where $t_L \equiv \eta L^3 /\kappa \sim
10^{8}$ s.} \label{fig:MSD0}
\end{figure}

Here, we now discuss the dynamical aspects of Eq.\
(\ref{Uneq_time_corr}). For simplicity, we assume that the curvature
coupling is small, {\em i.e.\ }$\omega_h \omega_\psi \gg \beta \gamma$,
which is indeed the case for the experiments in Ref.\ \cite{girard}.  However,
this approximation may not be always true in general and some of the conclusions
below may be modified in the strong coupling limit. Within the small curvature coupling approximation,
Eq.\ (\ref{Uneq_time_corr}) simplifies to \begin{equation}
\label{Granek_uneq_time}
\langle h({\bf q},t)h(-{\bf q},0) \rangle \simeq \frac{\Gamma_1  }{2 \,
\omega_h}\,e^{-\omega_h t} + \frac{\Gamma_2  \beta^2}{ 2 \,(\omega_h^2-\omega_\psi^2 )}
\left[ \frac{e^{-\omega_\psi t}}{\omega_\psi} - \frac{e^{-\omega_h t}}{\omega_h} \right].
\end{equation}
The central quantity of experimental interest is the MSD of a
membrane point defined by $\langle \Delta h^2(t) \rangle \equiv
\langle \left [\,h({\bf r}, t) - h({\bf r}, 0) \right
]^2\,\rangle= \int_{\pi/L}^{\pi/a} q\,(dq /2 \pi)\, [ \,\langle
h({\bf q},0)h(-{\bf q},0) \rangle  - \langle h({\bf q},t)h(-{\bf
q},0)\rangle \,]$, where $a$ and $L$ are, respectively, the
microscopic and the system size. They introduce two time scales
$t_a \sim \eta\,a^3 /\kappa$ and $t_L \sim \eta L^3 /\kappa$ that
are assumed to be, respectively, the shortest and longest time
scales in the problem.  Note also that $\langle \Delta h^2(t)
\rangle$ roughly corresponds to transverse fluctuations of a
particle embedded in a membrane in microrheological experiments
\cite{levine}. From Eq.\ (\ref{Granek_uneq_time}), we see that
there are two contributions: $\langle \Delta h^2(t) \rangle =
\langle \Delta h^2_{th}(t) \rangle + \langle \Delta h^2_{a}(t)
\rangle$. The first term comes from thermal fluctuations with the
well-known scaling law: $\langle \Delta h^2_{th}(t) \rangle=0.17
(k_B T/\kappa^{1/3} \eta^{2/3}) t^{2/3}$ \cite{Granek2}.  The
second term arises from the diffusion of the pumps. There are two
cases to consider: i) in the permeation dominated regime, in which
$\beta \simeq \lambda_p F_1$, we find\bb
\langle \Delta h^2_{a}(t) \rangle  \simeq \left \{    \begin{array}{ll}
                    1.62\,\frac{k_B T \Lambda \lambda_p^2 F_1^2 \eta^{4/3}}
  {\kappa^{4/3}} t^{5/3},  &\mbox{ for $\, t \ll t_{c1}$,}\\
                    \frac{k_B T \lambda_p^2 F_1^2}{6 \pi D \chi} t \ln
  \left( {t/t_{c1}} \right),  &\mbox{ for $\, t \gg t_{c1},$}
                    \end{array}   \right.
\label{Granek MSD}
\en
where $t_{c1} \equiv \kappa^2/(16 \eta^2 D^3)$. Thus, there is a superdiffusion at short time which
crosses over to an almost normal diffusion at long time.  This is
similar to the findings in Ref.\ \cite{Granek}, where the dynamics
of the PB model was analyzed. However, Eq.\
(\ref{Granek_uneq_time}) does not contain a term that arises from
the switching of the pumps that is explicitly included in the PB
model. Note also that the superdiffusion in Eq.\ (\ref{Granek
MSD}) is a purely nonequilibrium phenomenon in the sense that
membranes with passive inclusion only show subdiffusion
\cite{divet}. ii) In the experimental relevant regime in which $\lambda_p
=0$, {\em i.e.\ }permeation is negligible \cite{note1}, and $\beta=-(P_1 w+\Xi)/(4 \eta)\,q\equiv - \beta_1 q$,
we now find\bb \langle \Delta h^2_{a}(t) \rangle \simeq \left \{
\begin{array}{ll}
                    -0.85\,\frac{k_B T \beta_1^2 \eta^2 \Lambda}{\kappa^2}
                    t \ln \left( {t/t_{c1}} \right),  &\mbox{ for $t \ll t_{c1}$,}\\
                    1.08 \,\frac{k_B T \beta_1^2 \eta^{2/3}}
  {D \chi \kappa^{2/3}} t^{1/3},  &\mbox{ for $\, t \gg t_{c1}.$}
                    \end{array}   \right.
\label{My MSD}
\en
In contrast to case (i), the MSD shows an almost normal
diffusion at short time and a sub-diffuse regime at long times.
We have verified these scaling laws [Eqs.\ (\ref{Granek MSD}) and (\ref{My MSD})] by numerically
computing the MSD as shown in Fig.\ \ref{fig:MSD0}. For case (i), the thermal MSD
dominates at short time, while the active MSD dominates at long time. In constrast,
for case (ii), the thermal MSD is the dominant contribution.  Note also that Eq.\ (\ref{My MSD})
suggests that membranes containing active pumps and those containing passive pumps
obey the same scaling law.  Therefore, the RTP model leads to the conclusion
that dynamical measurements cannot distinguish
active membranes from passive ones in the experimentally relevant regime.
However, as noted above, the RTP model ignores the stochastic nature of the
pumps, which may be a serious approximation made in so far
as dynamics is concerned. The natural question to ask is: how does the shot noise
contribute to the membrane dynamics?  We address this question below.

\section{Active membranes with temporal force fluctuations}
\label{sec_fluctuations in active forces}

Recall that the active force density $F_A({\bf r}, t)$ contains
$F_1({\bf r}, t)$ and $F_2({\bf r}, t)$ that are assumed to be
constant in the RTP model.  Thus, the most direct way to
incorporate pump stochasticity is to allow them to fluctuate in
time.  Here, we focus on the direct force fluctuations: $F_1({\bf
r}, t) = F_1 + \delta F(t)$ in the $\lambda_p \sim 0$ regime.
Assuming separation of time scales, $\delta F(t)$ is described by
a Gaussian white noise with zero mean: $\langle \delta F(t)
\rangle =0$ and $\langle \delta F(t) \delta F(t') \rangle = 2 W
\delta(t-t') $, where $W$ characterizes the strength of the
fluctuations. This generalization of the RTP model may
capture the effects of fast temporal fluctuations of the
nonequilibrium forces exerted on the fluids by pumps. Incorporating $\delta F(t)$ into the RTP
equations, Eq.\ (\ref{Langevin1}), we see that it contributes an
additional term $\delta \beta(t)\,\psi({\bf q},t)$ to the
$h$-equation with $\langle \delta \beta(t)\delta \beta(t') \rangle
= \Gamma_s \delta(t-t')$, where $ \Gamma_s  \equiv 2 W
[\,\Omega(q) / (4 \eta q)]^2$ and $\Omega(q) = (1+ q w)\,e^{-q w}
- (1 + q w')\,e^{-q w'} $ is the ``structure factor" for the force
dipole \cite{bassereau2,sankararaman}.
\begin{figure}
{\par {{\includegraphics[scale=0.35]{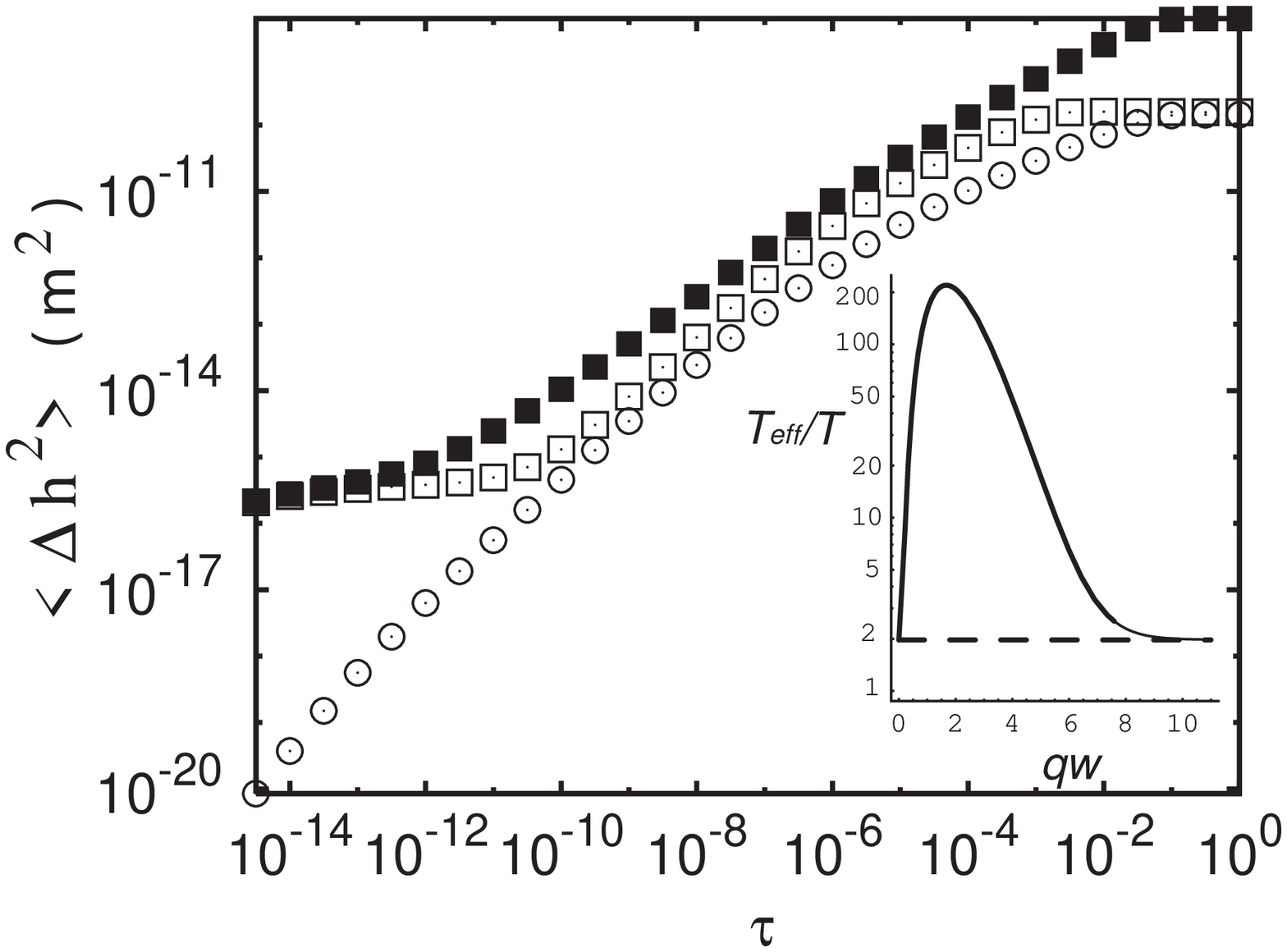}}
{\includegraphics[scale=0.35]{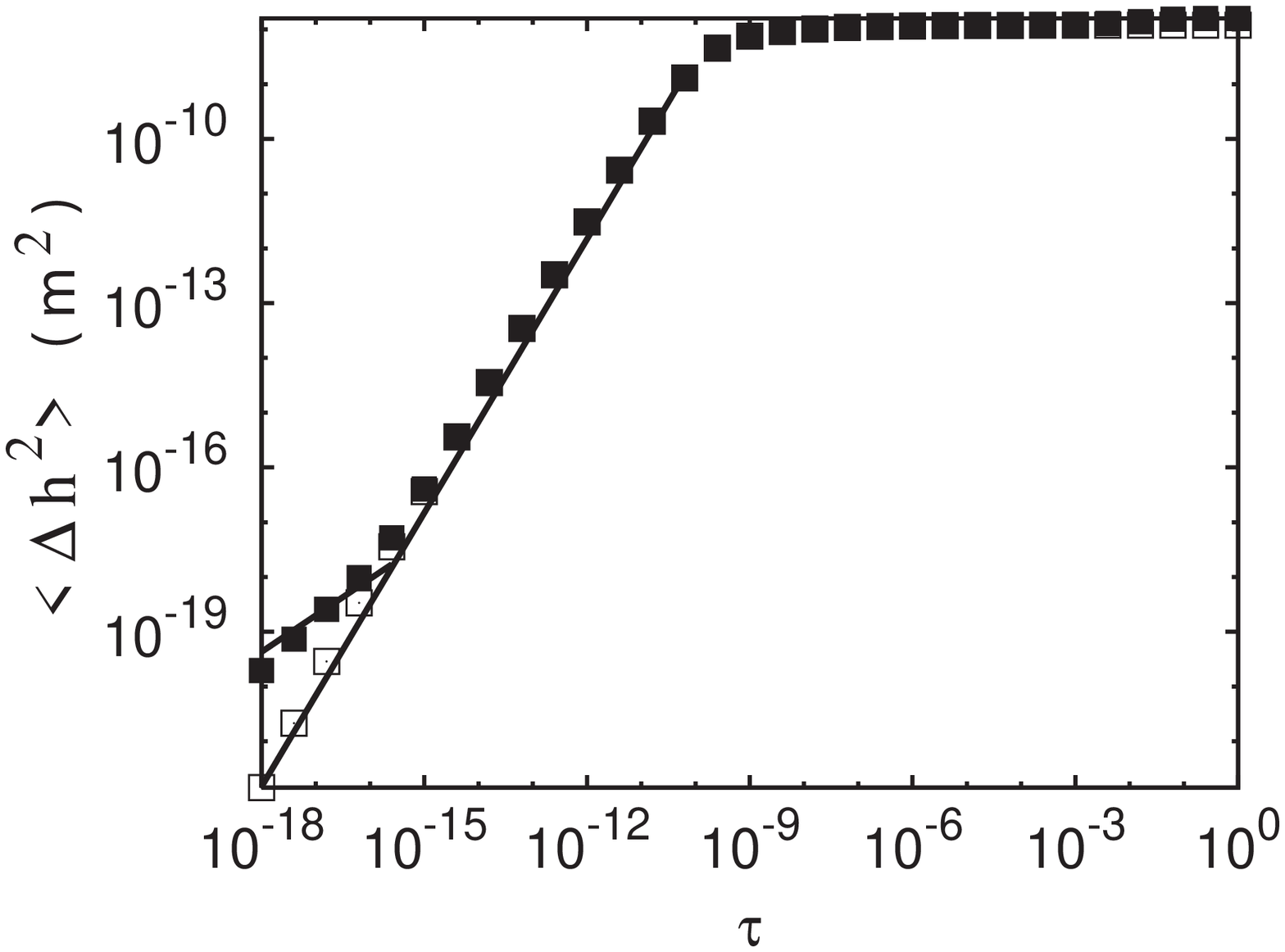}}}
\par}
\caption{The active part of the MSD $\langle \Delta h^2_a (t)
\rangle$ (open squares) and the overall MSD (solid squares) in the
presence of temporal force fluctuations as obtained from Eq.\
(\ref{FF}).  For short time, the behavior of the active MSD is
drastically different from that of the RTP model (open circles).
Inset: The scale dependent effective temperature
$T_{\mbox{\scriptsize eff}}(q)$ induced by force fluctuations
(solid curve). In the RTP model, $T_{\mbox{\scriptsize eff}}/T
\simeq 1.9$ (straight line) for parameters listed in Fig.\
\ref{fig:MSD0} and $W= 3.4 \cdot 10^{-26}\,$N$^2\,$s.
\label{fig:vp}} \caption{Same as Fig.\ \ref{fig:MSD0} but for the
two-state model as obtained from Eq.\ (\ref{2statecorr}). The
parameters are same as Fig.\ \ref{fig:MSD0}, and $k_a =
0.1\,$s$^{-1}$, $\tau_R \sim 10^{-2}\,$s, and
$\Gamma_3=10^{24}\,$s$^{-1}$m$^{-2}$. In contrast to the RTP model
in the $\lambda \sim 0$ limit, this MSD exhibits superdiffusion
which scales as $\sim t^{5/3}$.} \label{Fig:2state}
\end{figure}

The Langevin equations of this model involve multiplicative
and additive noises in a spatially extended system \cite{Sancho,Kirone},
which in general could pose a mathematically challenging problem; however,
in this case they can be solved exactly. It is straightforward to derive the
corresponding Fokker-Planck equation \cite{Risken}, which describes the time evolution of
the probability distribution ${\cal P}[\{h({\bf q})\}, \{\psi({\bf q})\}; t ]$.
Since different ${\bf q}$'s are decoupled,
we can write ${\cal P}[\{h({\bf q})\}, \{\psi({\bf q})\}; t ] =
\prod_{\bf q}\,P[h({\bf q}), \psi({\bf q}); t ]$, where $P[h({\bf q}),\psi({\bf q});t]$
satisfies \begin{equation}
\label{FP}
\partial_t P  =\partial_h \left[ \left( \omega_h h - \beta
\psi \right) P \right] + \partial_\psi
\left[ \left( \omega_\psi \psi - \gamma h \right) P \right]  + ({\Gamma_1 / 2})\,
\partial_h^2 P  + ({\Gamma_2 / 2})\,\partial_\psi^2 P  + ({\Gamma_s\,\psi^2 / 2})\,
\partial_h^2 P.
\end{equation}
The last term in Eq.\ (\ref{FP}), arising from force fluctuations,
is a nonlinear term which may render Eq.\ (\ref{FP}) difficult to
solve. However, we find, quite remarkably, a closed set of equations
for the moments of the form $\Psi_{m,k}(t) \equiv \langle
h^{m}(t)\,\psi^k(t) \rangle$, which directly follows from Eq.\
(\ref{FP}):\begin{eqnarray}
\partial_t \Psi_{m,k}(t) &=& -m\left (\,\omega_h \Psi_{m,k} - \beta \Psi_{m-1,k+1} \right ) -
k \left ( \omega_\psi \Psi_{m,k} - \gamma \Psi_{m+1,k-1} \right ) \nonumber \\
&+&  \,{m(m-1)} \left (\, \Gamma_1\, \Psi_{m-2,k}
+ \Gamma_s\, \Psi_{m-2,k+2 } \,\right )/2 + {\Gamma_2\, k(k-1)} \Psi_{m,k-2}/2,
\label{Eq for moments}
\end{eqnarray}
where $m= n- k$, $0 \leq k \leq n$, and $n$ is a positive integer.
From Eq.\ (\ref{Eq for moments}), we can compute all the moments of the steady-state
distribution which, in contrast to the RTP model, is clearly non-Gaussian.
Furthermore, we can obtain the exact {\em two-point} correlation function:\bb
\langle h({\bf q},t)h(-{\bf q},0) \rangle   =
{2\,\Gamma_1 \left[\,M_+\,e^{-\omega_+ t} - M_-\,e^{-\omega_- t}\,\right] \over
\left(4 A B -\Gamma_s \gamma^2 \right)\left(\omega_+ -\omega_- \right)}
   + { 2\, \Gamma_2\,\left[\,N_+\,e^{-\omega_- t} -  N_-\,e^{-\omega_+ t}\,\right] \over
   \left(4 A B -\Gamma_s \gamma^2 \right)\left(\omega_+ -\omega_- \right)},
\label{FF} \en where $N_{\pm} = \beta^2 \omega_{\pm} + \Gamma_s [
B \left( \omega_{\pm} - \omega_h \right) + \gamma \beta \,]/2$.
Note that Eq.\ (\ref{FF}) reduces to Eq.\ (\ref{Uneq_time_corr})
if we set $\Gamma_s=0$, as it should be.  It is interesting to
observe that the denominator in Eq.\ (\ref{FF}) contains the
factor $4 A B -\Gamma_s \gamma^2 $, which could become zero at
finite $q$ for sufficiently large $W$, signaling a dynamical
instability. An analysis of this instability will be presented in
a separate publication, along with the mathematical details
leading to Eq.\ (\ref{FF}). For small $W$, we find that $\langle
h({\bf q},0)h(-{\bf q},0) \rangle \simeq k_B T_{\mbox{\scriptsize
eff}}(q)/(\kappa_e q^4)$, with a scale dependent effective
temperature: $T_{\mbox{\scriptsize eff}}(q) = \kappa_e T [ 1 +  W
\Omega(q)^2  / (4 \eta q \chi) + P_1 w ( \Xi + P_1 w )/( \chi
\kappa )]/( \kappa_e - P_1 w\, \Xi/\chi)$, which is plotted in the
inset of Fig.~\ref{fig:vp}. It shows that force fluctuations
greatly enhance membrane fluctuations mostly in the region $q \sim
1/w$. Note also that $T_{\mbox{\scriptsize eff}}(q)$ depends on
the viscosity of the solvent, whereas $T_{\mbox{\scriptsize eff}}$
in the RTP model does not. Furthermore, temporal force
fluctuations contribute significantly to the active MSD, {\em
i.e.} the second term in Eq.\ (\ref{FF}), as can be seen in Fig.\
\ref{fig:vp}. Note that the active MSD in our model is drastically
increased at short time, though it does not predict
superdiffusion.  Thus, in contrast to the RTP model, the short
time behavior of the overall MSD is now dominated by the active
contributions.

\section{Modelling shot-noise with a two-state model}
\label{twostate}

A two-state model has recently been introduced to address the
dynamical instability of an active membrane containing inclusions
with two internal conformational states \cite{chen}.  Such a
two-state model of pumps switching between ``on" and ``off" states
may also capture the stochastic nature of the pumps.   It may be a
natural model for typical ion channels since they undergo random
transitions between open and closed states \cite{cell,Peter}. We
model the transition between ``on" and ``off" as a chemical
reaction with rate constants $k_p$ and $k_a$: $\mbox{on}
\stackrel{k_p}{\rightarrow} \mbox{off} \stackrel{k_a}{\rightarrow}
\mbox{on}$, and introduce imbalance fields for the ``on" pumps,
$\psi_a({\bf r},t) = \rho^{\uparrow}_{\mbox{\scriptsize on}}({\bf
r},t) - \rho^{\downarrow}_{\mbox{\scriptsize on}}({\bf r},t)$, and
``off" pumps, $\psi_p({\bf r},t) =
\rho^{\uparrow}_{\mbox{\scriptsize off}}({\bf r},t) -
\rho^{\downarrow}_{\mbox{\scriptsize off}}({\bf r},t)$.  The
active contribution to the force exerted on the fluid is assumed
to come only from the ``on" pumps: $\vec{F}_{A}({\bf x},t) = F_1\,
\psi_a({\bf r},t)\,\left[\,\delta(z-w)-\delta(z+w') \,
\right]\,\hat{\bf z}$. In analogy to Eq.\ (\ref{Langevin1}), the
general equations of motion for the two-state model in the regime
with negligible permeation ($\lambda_p =0$) are\bb
\begin{array}{l}
\partial_t h({\bf q},t) + \omega_h h({\bf q},t) =
\beta_a \psi_a({\bf q},t) + \beta_p \psi_p({\bf q},t) + \mu({\bf q},t),\\
\partial_t \psi_a({\bf q},t)  + \omega_{a} \psi_a({\bf q},t)  = \gamma_{a} \,h({\bf q},t)
 + k_a  \psi_p({\bf q},t) + \nu_a({\bf q},t)+ \xi({\bf q},t), \\
\partial_t \psi_p({\bf q},t)  + \omega_{p} \psi_p({\bf q},t) = \gamma_{p} \,h({\bf q},t)
+ k_p \psi_a({\bf q},t)+ \nu_p({\bf q},t) - \xi({\bf q},t),
\end{array}
\label{Langevin2}
\en
where $\omega_h = \kappa q^3/(4 \eta)$, $\beta_a = - (P_1 w + \Xi_a )q/ (4 \eta) \equiv - \beta_1 q$,
$\beta_p = - \Xi_p q/ (4 \eta)$, $\omega_{a,p} = D_{a,p} q^2 + k_{p,a}$,
and $\gamma_{a,p} = - \Lambda_{a,p} \Xi_{a,p} q^4$. As
in the RTP model, $\mu({\bf q},t)$, $\nu_a({\bf q},t)$, and $\nu_p({\bf q},t)$
are thermal noises with zero mean and $\langle \mu({\bf q},t)\,\mu(-{\bf q},t')
\rangle = k_B T /(2\eta q )\delta(t-t')$ and $\langle \nu_i({\bf q},t)\,\nu_j(-{\bf q},t') \rangle =
2 k_B T \Lambda_i\,q^2 \delta_{ij}\,\delta(t-t')$. In Eq.\ (\ref{Langevin2}),
$\xi({\bf r},t)$ denotes the ``chemical noise" associated with
the transitions between the two states; since the switching process may
involve ATP, which is clearly not an equilibrium process, its variance
is not constrained by the Fluctuation-Dissipation Theorem.  We simply assume
that $\xi({\bf r},t)$ has zero mean and $\langle\,\xi({\bf r},t) \xi({\bf r}', t') \rangle
= \Gamma_3\,\delta({\bf r} -{\bf r}')\delta(t-t')$, with $\Gamma_3$ being constant
in space and time.

Although an exact, but complicated, expression for the two-point correlation function
can be obtained analytically, we find it convenient to introduce the following
simplifying approximations in order to illustrate the essential features:
$\Lambda_{a} = \Lambda_{p} = \Lambda$, $\chi_{a} = \chi_{p} = \chi$,
$D_{a} = D_{p} = D$, $ \beta_a^2 \gg \beta_p^2$, and $\omega_h \omega_a \gg \beta_{a} \gamma_a$.
Under these approximations, we obtain\begin{eqnarray}
\langle h({\bf q}, t)h(-{\bf q}, 0)\rangle &=& \frac{
\Gamma_1}{2\,\omega_h} e^{-\omega_h t} + \frac{\Gamma_2  \beta_a^2 k_a \omega_p}{
\left(\omega_h^2-\omega_1^2 \right)\left(\omega_2^2-\omega_1^2 \right)} \left[
\frac{e^{-\omega_1 t}}{\omega_1} - \frac{e^{-\omega_h
t}}{\omega_h} \right] \nonumber \\
&+& \frac{ \beta_a^2}{ 2 \left(\omega_h^2-{\omega}_2^2 \right)} \left [ \Gamma_3 + \Gamma_2
\left ( 1 - { 2 k_a \omega_p \over \omega_2^2 - \omega_1^2 }\right ) \right ]\left[
\frac{e^{-{\omega_2} t}}{\omega_2} - \frac{e^{-\omega_h t}}{\omega_h} \right],
\label{2statecorr}
\end{eqnarray}
where $\Gamma_1 \equiv k_B T /(2 \eta q )$, $\Gamma_2 \equiv 2 k_B
T \Lambda\,q^2$, $\omega_1 = Dq^2$, $\omega_2 = D q^2 + 1/\tau_R$,
and $\tau_R \equiv 1/( k_a + k_p)$. The first two terms in Eq.\
(\ref{2statecorr}) have the same physical origins as the
corresponding terms in Eq.\ (\ref{Granek_uneq_time}). In
particular, the second term gives rise to a MSD which has the same
scaling laws as in Eq.\ (\ref{My MSD}), except that its magnitude
is reduced by a factor of $\sim (k_a \tau_R)^2 $. The third term,
which is absent in the RTP model, arises from the noise associated
with the switching process between ``on" and ``off" states of the
pumps and their diffusion.  It is analogous to the shot-noise term
in the PB model \cite{Granek} but ours contains curvature
couplings, which are absent in the PB model.  Note also that the
analysis of Ref.\ \cite{Granek} assumes that the system is in the
permeation dominated regime \cite{note1}. Assuming $\Gamma_3$ is
sufficiently large, the MSD arising from the third term in Eq.\
(\ref{2statecorr}), $\langle \Delta h^2_{s}(t) \rangle$, has to
the following scaling laws\bb
\langle \Delta h^2_{s}(t) \rangle
\simeq \left \{    \begin{array}{ll}
                    0.32\,\Gamma_3  \beta_1^2 (\eta /\kappa)^{4/3} t^{5/3},
                    &\mbox{ for $t \ll \tau_R$,}\\
                    0.16 \,\Gamma_3  \beta_1^2 (\eta /\kappa)^{4/3}\tau_R^{5/3},
                    &\mbox{ for $ t \gg \tau_R.$}
                    \end{array}   \right.
\label{shotnoiseMSD}
\en
Therefore, we find that the MSD exhibits
superdiffusion $\sim t^{5/3}$ at short time.  This is
qualitatively different from the prediction of RTP model in Eq.\
(\ref{My MSD}) as well as the force fluctuations model in Eq.\
(\ref{FF}). Since the first two terms in Eq.\ (\ref{2statecorr})
exhibit only subdiffusion at short time, $\langle \Delta
h^2_{s}(t) \rangle$ is the dominant contribution to the MSD, which
is plotted in Fig.\ \ref{Fig:2state}.  As an estimate, $\tau_R$
may be about $0.01\,$s for typical ion pumps \cite{cell,time}.

In summary, we have
generalized the RTP model by incorporating the shot noise of the
pumps and demonstrated its significance to the dynamics of an
active membrane.

\acknowledgments
We acknowledge many fruitful discussions with P.\ Bassereau, P.\ Girard,
R.D.\ Kamien, T.C.\ Lubensky, K.\ Mallick, K.\ Sekimoto, J.\ P\'ecr\'eaux,
and J.\ Prost.  A.W.C.L.\ acknowledges support from the NSF through the
MRSEC Grant DMR-0079909 and the CNRS-NSF Binational Grant INT99-10017.


\begin{thebibliography}{0}
\bibitem{flat}
We assume that the deformation of the membrane is small, so that
in a linearized treatment the normal to the membrane coincides with the $z$ direction in the lab frame.
\bibitem{note1}
As disscussed in Ref.\ \cite{bassereau2}, the permeation contributions become
important only on length scales $\gg 1\,\mbox{cm}$.  Thus, only the
$\lambda_p  \sim 0$ limit is realized in experiments.
\end{thebibliography}
\end{document}